\documentstyle[12pt,twoside,fleqn,espcrc1,graphicx]{article}
\title{Dynamics of Particle Production in Relativistic Nuclear Collisions}

\author{K. Tuominen\address{NORDITA, Blegdamsvej 17, 2100 Copenhagen
\O, Denmark}}
       
\begin{document}

\maketitle

\begin{abstract}
Saturation models for particle production in relativistic nuclear
collisions are discussed. In particular, I show that the predictions
from the high density QCD for the qualitative shape of $dN/dy$ are
very sensitive to the form of the unintegrated gluon distribution.

\end{abstract}

\section{Introduction}
After first years of running, RHIC has provided a lot of interesting
data \cite{RHICdata}.  The very basic observable, the number of
particles at central unit of rapidity, seems to indicate that
saturation models describe well the center of mass energy dependence
as well as the centrality dependence. The models I consider in this
talk are: The pQCD+saturation -model \cite{ekrt} and the high density
QCD calculation by Kharzeev and Levin (KL) \cite{KL}.  Although both
models agree with data and each other at central rapidity, they show
qualitative differences away from midrapidity.  We will trace the
origin of this discrepancy and see how it could be improved upon and,
along the way, we shall briefly remark how these two models would
predict the second important observable, the transverse energy.

\section{Models and results}
In the pQCD+saturation model the multiplicity of initially produced
gluons is evaluated under the assumption of collinear factorization
and including all quanta above the saturation scale $p_{s}$ which is
obtained as a self consistent solution of the saturation condition
\begin{equation}
\frac{dN}{dy}=T_{AA}\sum_{ijkl=q\bar{q}g}\int_{p_{s}}dp_\perp^2dy_2
x_1f_i(x_1,p_\perp^2)
x_2f_j(x_2,p_\perp^2)\frac{d\hat{\sigma}^{ij\rightarrow
kl}}{d\hat{t}}=p_{s}^2R_A^2.
\label{satcond}
\end{equation}
After solving for $p_{\rm s}$ one obtains the scaling laws \cite{ekrt}
\begin{eqnarray}
\left\{\begin{array}{l}
\frac{dN_g^{AA}}{dy}={\scriptstyle{1.38A^{0.92}\sqrt s^{0.38}}} \\
\frac{dE_{\perp,\mbox{\scriptsize{ini}}}^{AA}}{dy}=
{\scriptstyle{0.386A^{1.05}\sqrt s^{0.60}~\mbox{\small{GeV}}}}
\end{array} \right. 
\stackrel{\mbox{{\scriptsize{1D ideal}}}}{\mbox{{\scriptsize{expansion:}}}}
\left\{\begin{array}{l}
\frac{dN_\pi^{AA}}{dy}\approx\frac{dN_g^{AA}}{dy} \\
\frac{dE_{\perp,\mbox{\scriptsize{fin}}}^{AA}}{dy}
\approx(\frac{T_c}{T_{\mbox{\scriptsize{ini}}}})
\frac{dE_{\perp,\mbox{\scriptsize{ini}}}}{dy}
={\scriptstyle{3.48T_cA^{0.92}\sqrt s^{0.40}}}
\end{array} \right.
\label{pqcdsatpreds}
\end{eqnarray}
for the initial gluon multiplicity and transverse energy at $y=0$ and,
assuming ideal $1D$ expansion, for the multiplicity and transverse energy
of hadrons at $y=0$. In (\ref{pqcdsatpreds}) $T_c$ is 180 MeV and
evolution in the hadronic phase has been neglected as
this is compensated for by the development of the flow.

\begin{figure}[htb]
\begin{minipage}[t]{80mm}
\includegraphics[width=7cm,trim=0 100 0 100]{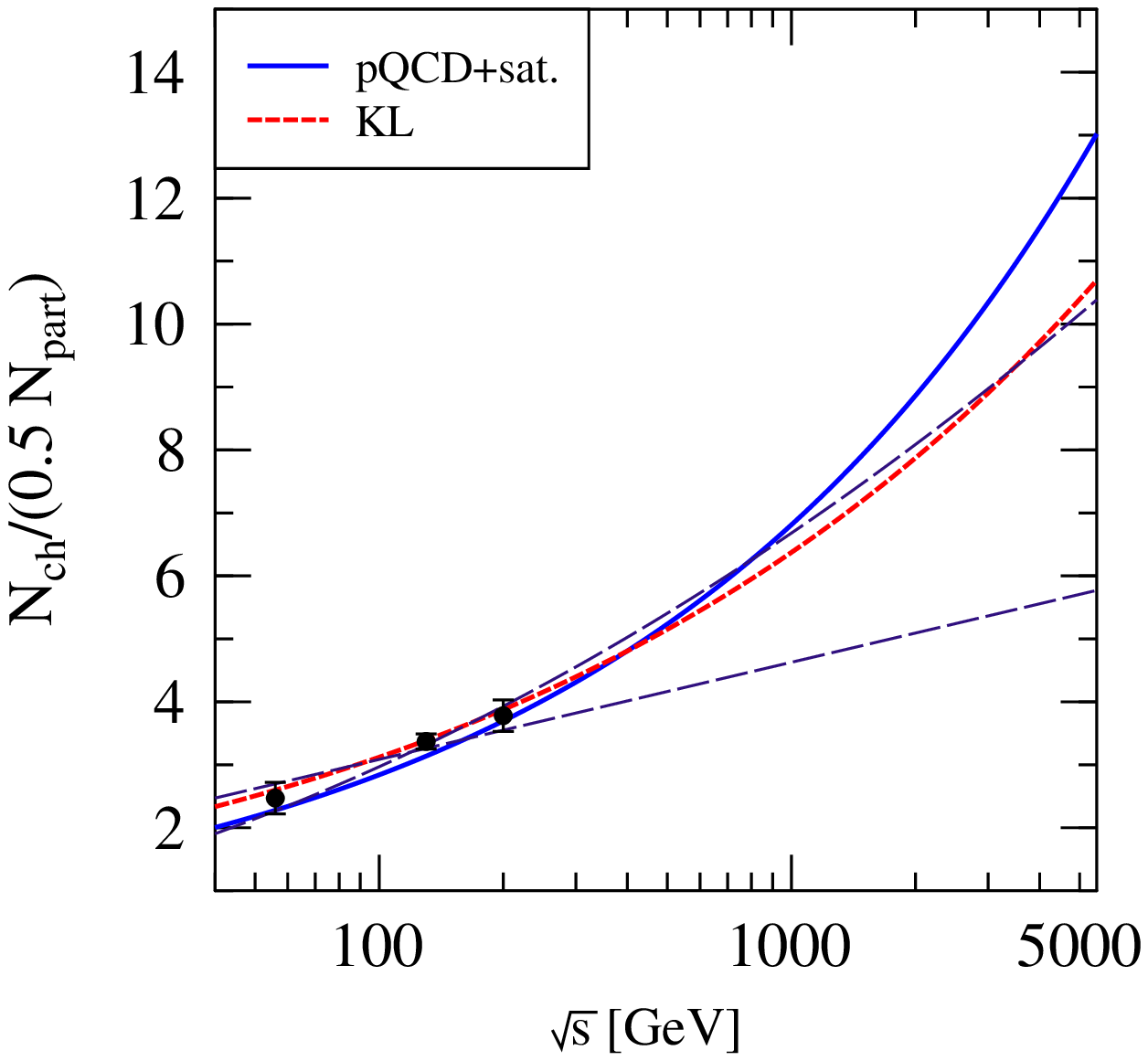}
\caption{Multiplicity in the central unit of pseudorapidity. Data is
from PHOBOS and dashed lines show the $\log(s)$ (straight) and
$\log^2(s)$ (curved) growths.}
\label{nfig}
\end{minipage}
\hspace{\fill}
\begin{minipage}[t]{75mm}
\includegraphics[width=5.7cm,trim=0 150 0 100]{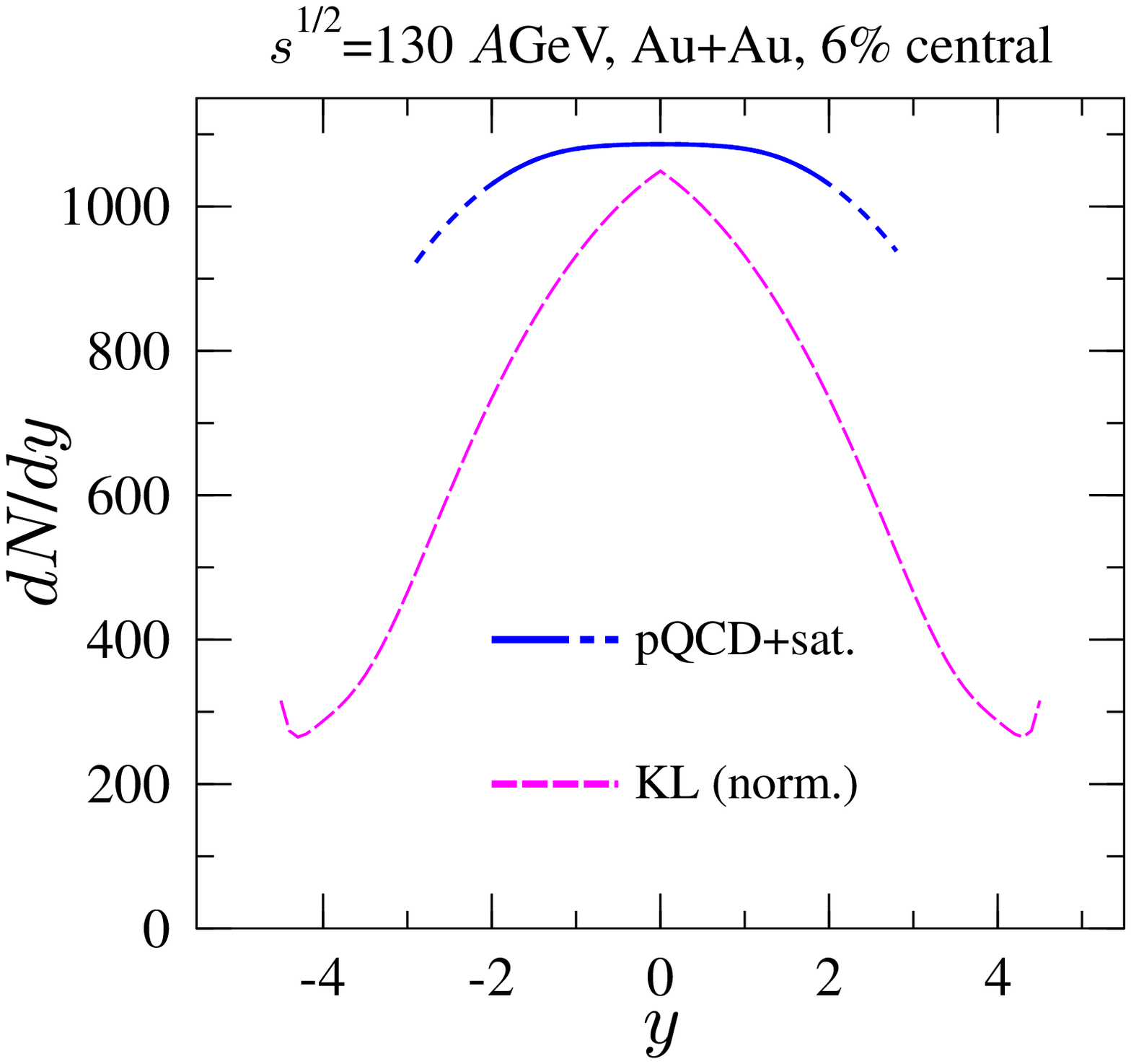}
\caption{$dN/dy$ from the two saturation models. Normalization of the
KL curve is such that after transforming, with an overall Jacobian
factor, from $y$ to $\eta$ it agrees with data. Figure is from \cite{ekrt2}.}
\label{dndy}
\end{minipage}
\end{figure}

The KL calculation also predicts powerlike growth of the number of particles
in central rapidity with $\sqrt s$ \cite{KL}:
\begin{equation}
N\sim N_{\rm part}(\frac{\sqrt s}{\sqrt s_0})^\lambda
\ln(\frac{Q_{\rm s}^2(\sqrt s_0)}{\Lambda^2_{\rm QCD}}
(\frac{\sqrt s}{\sqrt s_0})^\lambda),
\end{equation}
where $\lambda=0.25$ is related to the small Bjorken-$x$ growth of the
gluon structure function. The power is slightly smaller than the one
in the pQCD+saturation calculation.  The figure \ref{nfig} shows the
results from these two models and one sees that in the presently
available energy range the models are indistinguishable at central
(pseudo)rapidity. Extrapolation to LHC energies leads to a wider range
of predictions and not all models are distinguishable even there. For
the pQCD+saturation curve an effective value of 178 for $A$ was used,
as this corresponds to the 6\% centrality cut of the data \cite{errt}.
\vspace{1cm}
\vspace{-0.5cm} A crude estimate of $E_\perp$ in the
pQCD+saturation model can be obtained from Eq.(\ref{pqcdsatpreds}):
at $\sqrt s=130$ GeV $E_\perp=520$ GeV and at $\sqrt s=200$ GeV
$E_\perp=620$ GeV. Transverse expansion effects increase the
above numbers a little \cite{errt}.  PHENIX data at
$\sqrt s=130$ is $E_\perp=578$ GeV \cite{PHENIXet}.

The KL calculation has been shown to reproduce pseudorapidity
distributions of hadrons at $\sqrt s=130$ GeV well \cite{KL}. From the
theoretical point of view it is more instructive to look at the
rapidity distributions of initial gluons, see Fig. \ref{dndy}. The
pQCD+saturation model leads to a broad gaussian which, after
transforming to pseudorapidity and comparing with data, overshoots at
large rapidities.  If saturation dynamics lead to flat $dN/dy$ near
$y\sim 0$, this behaviour has to change around $y\sim y_{\rm{beam}}/2$
and the saturation dynamics must be replaced by some other
fragmentation region dynamics. The KL result is qualitatively very
different: it has a discontinuity in the first derivative of $dN/dy$
at $y=0$ and is exponentially suppressed away from it.

\subsection{High Density QCD methods}
The result of KL is based on the GLR equation \cite{glr}:
\begin{equation}
\frac{dN}{dy}=\frac{1}{\sigma_{\rm{in}}}\int dp_\perp^2\frac{\alpha}{p_\perp^2}
\int^{p_\perp^2} dk_\perp^2\phi_A(x_1,k_\perp^2)\phi_A(x_2,(p-k)_\perp^2), 
\label{glrequ}
\end{equation}
which requires an {\em{ansatz}} for $\phi_A(x,k_\perp^2)$. The ansatz used
by KL is effectively
\begin{equation}
\phi_A(x,k_\perp^2)=\frac{2}{3\pi^2}\frac{S_A}{\alpha}
[\theta(Q_{s,A}^2-k_\perp^2)
+\epsilon
\frac{Q_{s,A}^2}{k_\perp^2}\theta(k_\perp^2-Q_{s,A})],
\label{toy}
\end{equation}
where parametrically $\epsilon\sim O(\alpha^2)$ and sets the
relative normalization of the saturated and perturbative parts of the
unintegrated gluon distribution. KL choose $\epsilon=0$ and regard the
tail as a small correction.  With ansatz (\ref{toy}) one obtains to
leading logarithmic accuracy
\begin{equation}
\nonumber \frac{dN}{dy}= \frac{2}{3\pi^2}\frac{S_A Q_{s,A}^2e^{-\lambda
|y|}}{\alpha(Q_{s,A}^2)}[2+2\lambda
y(1+\epsilon)+2\epsilon[(1-\lambda)+\epsilon(1-\frac{3}{2}\lambda+\lambda
y)].
\end{equation}
Setting $\epsilon=0$ the KL result is reproduced. With $\epsilon=1$
and $\lambda=0.25$ one sees that the term $2\epsilon[\dots]$
originating entirely from the tail of the distribution and the first
term from saturation region are roughly equal. For the transverse
energy the tail is more important:
\begin{equation}
\frac{dE_\perp}{dy}=\frac{2}{3\pi^2}\frac{S_A Q_{s,A}^3e^{-3\lambda
|y|/2}}{\alpha(Q_{s,A}^2)}[\frac{4}{3}+2\lambda y(1+\epsilon)
+4\epsilon[(1-2\lambda)+2\epsilon(1-\frac{7}{2}\lambda+\frac{\lambda}{2}
y]].\end{equation} Of course it is plausible that the relative
normalization is not given by $\epsilon=1$ but by some smaller value,
since $\epsilon\sim O(\alpha^2)$, which would validate the exclusion
of the tail.

\subsection{Reshaping $dN/dy$}
\begin{figure}[htb]
\begin{minipage}[t]{80mm}
\includegraphics[width=7cm,trim=0 150 0 0]{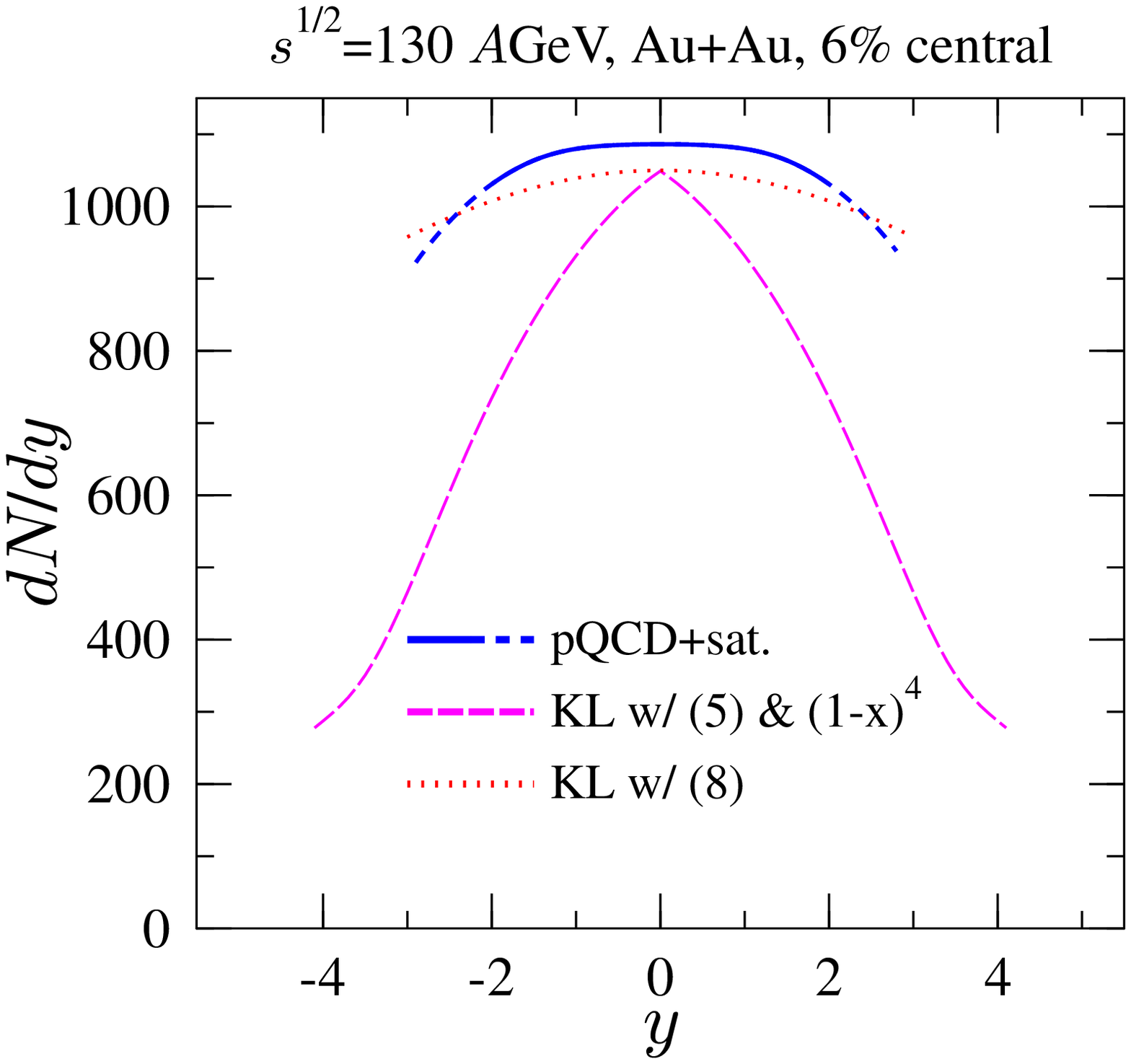}
\caption{$dN/dy$ from the different calculations discussed in the
text.}
\label{dndynew}
\end{minipage}
\hspace{\fill}
\begin{minipage}[t]{75mm}
\includegraphics[width=5.7cm,trim=0 150 0 0]{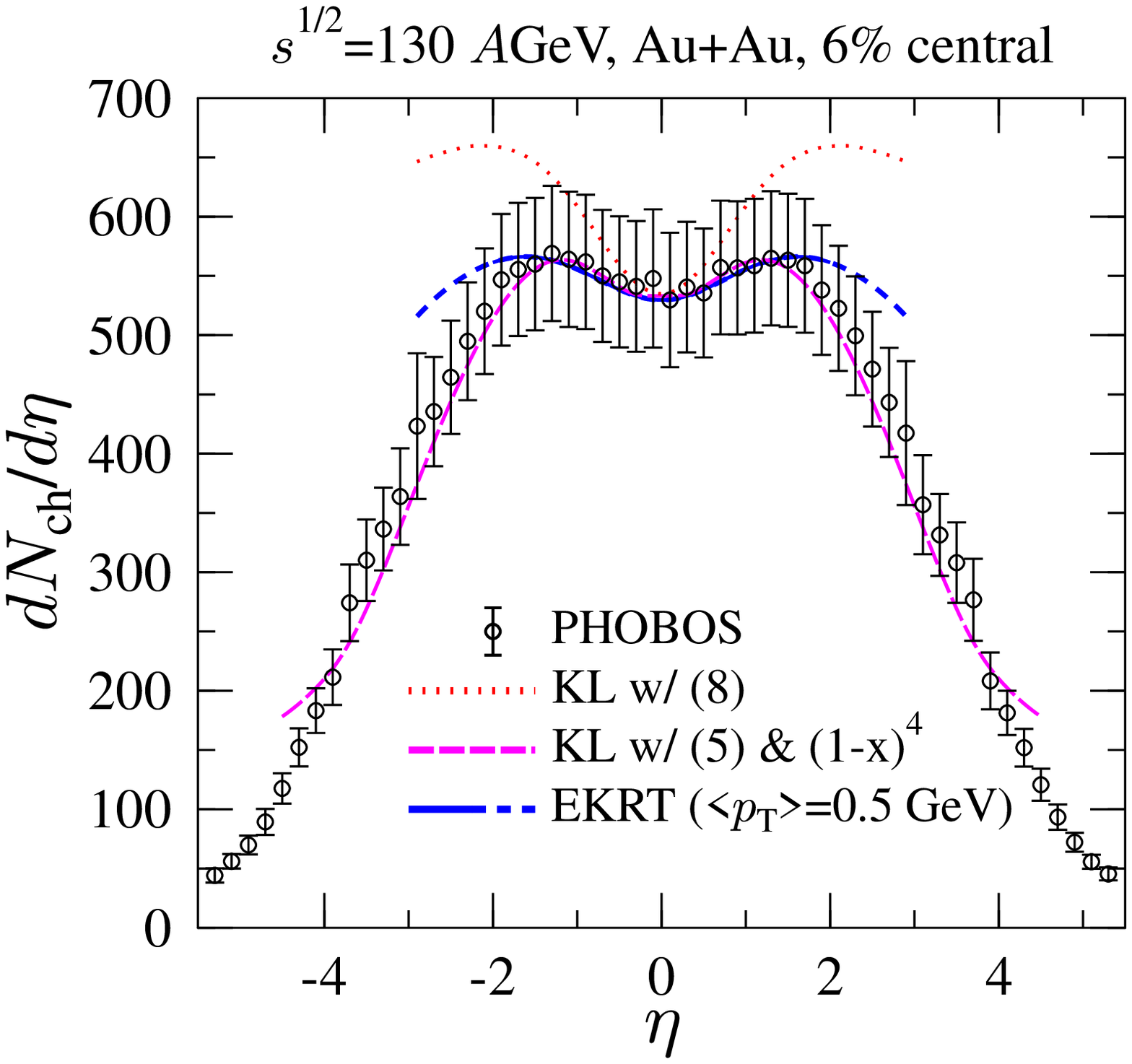}
\caption{$dN/d\eta$ from the different calculations discussed in the text.}   
\label{dndeta}
\end{minipage}
\vspace{-0.5cm}
\end{figure}

The ansatz (\ref{toy}) is probably too simple and one should try
for example the one from \cite{iancu}:
\begin{equation}
\phi_A(x,k_\perp^2)\sim\int\frac{d^2z}{\alpha z^2}e^{-i{\bf{k\cdot z}}}
(1-e^{-\frac{1}{4}Q_{s,A}^2(x)z^2\ln(z_0^2/z^2+1)}).
\label{realphi}
\end{equation}
Using this in (\ref{glrequ}) one finds that the form of $dN/dy$
is
\begin{equation}
\frac{dN}{dy}\sim \frac{S_A Q_{s,A}^2}{\alpha(Q_{s,A})}\frac{\lambda y}{\sinh(\lambda y)}
[6+\frac{1}{18}(2\lambda y)^2-\frac{1}{1800}(2\lambda y)^4+\dots].
\label{realn}
\end{equation}
This is very different from the one obtained with ansatz (\ref{toy})
and is closer to a broad gaussian as can be seen from
the figure \ref{dndynew}, which shows all of the discussed multiplicity
distributions.

One needs also to take into account the large $x$ behaviour of the
gluon distribution, $xG\sim(1-x)^4$, not contained in Eq.(\ref{realphi});
whether this is sufficient to make agreement with data remains unclear
at the moment. However near $y\sim 0$, one expects result
(\ref{realn}) to be dominated by small $x$ part of $\phi_A$, and one
might try to transform to pseudorapidity and compare with data. For KL
this is done by an overall factor and for pQCD+saturation -model by
assuming exponential $p_\perp$-spectra for pions. For details, see
\cite{KL,ekrt2}. From Fig. \ref{dndeta} one sees that the
transformation of (\ref{realn}) with the overall factor leads to a
very large dip around the central pseudorapidity. Hence, one should
rather use the $p_\perp$-distributions to carry out the
transformation. Probably the effects of the large $x$ behaviour of the
gluon distribution should be included already at $y\sim 0$, too.

\section{Conclusions}
Particle production over the whole experimentally accessible rapidity
range has been investigated using saturation models.  While these
models lead to very similar results at central unit of rapidity, they
seem to differ at nonzero rapidities. This difference was suggested to
originate from the choice for the unintegrated gluon structure
function in the KL calculation, and a different ansatz was shown to
lead to a gaussianlike distribution with a width comparable to the
pQCD+saturation -model result. The quality of the approximations such
as the neglect of the tail of the distributions and transformation
from $y$ to $\eta$ with an overall Jacobian factor, was shown to be
model dependent.

{\bf{Acknowledgements:}} I would like to thank K.J. Eskola, K. Kajantie
and P.V. Ruuskanen for collaboration and M. Gyulassy and D. Kharzeev
for discussions.


\begin{thebibliography}{9}
\bibitem{RHICdata}
K.~Adcox {\it et al.},  
Phys.\ Rev.\ Lett.\  {\bf 86}, 3500 (2001);
C.~Adler {\it et al.},  
Phys.\ Rev.\ Lett.\  {\bf 87}, 112303 (2001);
B.B.\ Back {\it et al.},  
Phys.\ Rev.\ Lett.\  {\bf 87}, 102303 (2001),
Phys.\ Rev.\ Lett.\  {\bf 88}, 022302 (2002);
I.~G.~Bearden {\it et al.},  
Phys.\ Lett.\ B {\bf 523}, 227 (2001),
Phys.\ Rev.\ Lett.\  {\bf 88}, 202301 (2002).

\bibitem{ekrt}
K.~J.~Eskola, K.~Kajantie, P.~V.~Ruuskanen and K.~Tuominen,
Nucl.\ Phys.\ B {\bf 570} (2000) 379.

\bibitem{KL}
D.~Kharzeev and E.~Levin,
Phys.\ Lett.\ B {\bf 523} (2001) 79.

\bibitem{errt}
K.~J.~Eskola, P.~V.~Ruuskanen, S.~S.~R\"as\"anen and K.~Tuominen,
Nucl.\ Phys.\ A {\bf 696} (2001) 715.

\bibitem{ekrt2}
K.~J.~Eskola, K.~Kajantie, P.~V.~Ruuskanen and K.~Tuominen,
hep-ph/0204034.


\bibitem{PHENIXet}
K.~Adcox {\it et al.}  
Phys.\ Rev.\ Lett.\  {\bf 87} (2001) 052301.

\bibitem{glr}
L.~V.~Gribov, E.~M.~Levin and M.~G.~Ryskin,
Phys.\ Rept.\  {\bf 100} (1983) 1.

\bibitem{iancu}
E.~Iancu, A.~Leonidov and L.~McLerran,
hep-ph/0202270.

\end{thebibliography}
\end{document}